\documentclass{aa}  
\usepackage{graphicx}
\usepackage{txfonts}
\usepackage[colorlinks,citecolor=blue,urlcolor=blue,filecolor=blue,linkcolor=blue]{hyperref}
\usepackage{amsmath}
\usepackage{amssymb}
\usepackage{upgreek}
\usepackage{natbib}
\usepackage{xcolor}

\newcommand{\msun}{\, {\rm M}_\odot}

\begin{document} 
    \title{The Crab Nebula at subarcsecond resolution with the International LOFAR Telescope}
   
    \author{M.\ Arias\inst{1,2}
        \and
        R.\ Timmerman\inst{3,4}
        \and 
        F. Sweijen\inst{3,4}
        \and
        R.\ J.\ van Weeren\inst{5}
        \and
        C.\ G.\ Bassa\inst{1}
    }

    \institute{
        ASTRON Netherlands Institute for Radio Astronomy, Oude Hoogeveensedijk 4, 7991\,PD Dwingeloo, The Netherlands\\
        \email{arias@astron.nl}  
        \and
        Anton Pannekoek Institute, University of Amsterdam, Science Park 904, 1098 XH Amsterdam, The Netherlands
        \and
        Centre for Extragalactic Astronomy, Department of Physics, Durham University, Durham, DH1 3LE, UK
        \and
        Institute for Computational Cosmology, Department of Physics, Durham University, South Road, Durham DH1 3LE, UK
        \and
        Leiden Observatory, Leiden University, PO Box 9513, 2300 RA Leiden, The Netherlands
    }

    \date{Received \today; Accepted }

    \abstract{
    We present International LOw Frequency ARray (LOFAR) Telescope (ILT) observations of the Crab Nebula, the remnant of a core-collapse supernova explosion observed by astronomers in 1054. The field of the Crab Nebula was observed between 120 and 168~MHz as part of the LOFAR Two Meter Sky Survey (LoTSS), and the data were re-processed to include the LOFAR international stations to create a high angular resolution ($0.43'' \times 0.28''$) map at a central frequency of 145~MHz. Combining the ILT map with archival centimeter-range observations of the Nebula with the Very Large Array (VLA) and LOFAR data at 54~MHz, we become sensitive to the effects of free-free absorption against the synchrotron emission of the pulsar wind nebula. This absorption is caused by the ionised filaments visible in optical and infrared data of the Crab Nebula, which are the result of the pulsar wind nebula expanding into the denser stellar ejecta that surrounds it and forming Rayleigh-Taylor fingers.
    The LOFAR observations are sensitive to two components of these filaments: their dense cores, which show electron densities of $\gtrsim1,000$~cm$^{-3}$, and the diffuse envelopes, with electron densities of $\sim50-250$~cm$^{-3}$. The denser structures have widths of $\sim0.03$~pc, whereas the diffuse component is at one point as large as $0.2$~pc. The morphology of the two components is not always the same. These finding suggests that the layered temperature, density, and ionisation structure of the Crab optical filaments extends to larger scales than previously considered.
    }


    \maketitle
    
\section{Introduction}

The Crab Nebula is one of the most thoroughly observed objects in the Galaxy \citep{hester08}. Its many names are a testament to its observational draw: it is called SN~1054 from the year it was observed as an optical transient by historical astronomers, Messier 1 for its optical nebula, 3C~144 and Taurus A for its radio nebula, and has the supernova remnant catalog name of G~$184.6-5.8$. It is also one of the brightest sources in the radio sky, with a flux density at 1~GHz of $1042\pm75$~Jy \citep{baars77}. 

The Crab Nebula consists of the Crab Pulsar (PSR B0531+21, the neutron star left behind in the supernova explosion), its synchrotron-emitting pulsar wind nebula, and a bright shell of thermally emitting stellar ejecta and swept-up interstellar material \citep{hester08}. It might also have a faint, undetected supernova remnant shell that surrounds the other components \citep{sollerman00}. The pulsar wind nebula is confined by, and pushing into, the thermally-emitting material; this results in Rayleigh-Taylor instabilities that give the appearance of a \lq cage\rq\ of thermal filaments enclosing the synchrotron nebula \citep{chevalier75,hester96}. The filaments are photoionised due to the radiation emitted by the synchrotron nebula, and show a variety of ionisation states \cite[e.g.][]{graham90,blair92} in elements typically present in supernova ejecta, such as oxygen, sulfur, iron, and nickel \citep{temim24}. Throughout this work, we refer to the \lq pulsar wind nebula\rq\ as the wind of relativistic particles powered by the loss of rotational energy of the pulsar, and to the \lq supernova remnant\rq\ as the expelled stellar ejecta and swept up circumstellar and interstellar media, including the supernova blast wave that would be seen as an (as of now undetected) shell surrounding the complete structure that we call the Crab Nebula. 

The ionisation structure of the ejecta filaments that enclose the pulsar wind nebula is layered, with lower ionisation states found in compact, sheltered cores surrounded by higher ionisation envelopes. The filament cores also contain dust, which can be seen as extinction features against the synchrotron nebula \citep{sankrit98,owen15}. This geometry --an extended synchrotron source with ionised material in the foreground-- can result in free-free absorption of the synchrotron emission at low ($<100~$MHz) radio frequencies \citep{arias18,arias19}, provided that the foreground material is sufficiently dense and/or cold.

The Crab Nebula has been observed extensively in the radio, in particular with the Karl G. Jansky Very Large Array (VLA) in all its configurations and at a variety of observing frequencies \citep[]{bietenholz90,kassim93,bietenholz04,bietenholz15}. Its integrated radio spectral index is constant at $\alpha=-
0.27\pm0.04$ (for $S_\nu \propto \nu^{\alpha}$) from 10~MHz to $\sim10^4$~GHz \cite[corresponding to the frequency of the synchrotron break;][]{bietenholz97}.

The \lq historical' VLA, the array as it was before the upgrade that finished in 2011, had two receivers at frequencies below 1~GHz: one at 74~MHz and one at 327~MHz. Combining observations at these frequencies with data at 1.5~GHz and 4.9~GHz, \cite{bietenholz97} carried out a radio spectral index study of the source with the aim of looking for spatial variations. The authors found that the spectral index of the nebula is spatially uniform between the 327~MHz, 1.4~GHz, and 4.9~GHz maps, but that at 74~MHz there is an unresolved absorption feature, which they ascribe to free-free absorption from the brightest optical knot with negative radial velocity. Here we confirm and expand on this interpretation with high-resolution, low-frequency imaging of the Crab Nebula.

The LOw Frequency ARray \cite[LOFAR,][]{vanhaarlem13} is a radio interferometer with a dense core located in the Netherlands, and individual stations spread further over the Netherlands and Europe. When the Dutch LOFAR stations are combined with the remaining LOFAR international stations the array is known as the International LOFAR Telescope (ILT), and provides sub-arcsecond resolution with its High Band Antennas (HBAs) centered at 150~MHz. In this work, we present low-frequency radio observations of the Crab Nebula with the ILT, where absorption effects due to the thermal filament cage surrounding the nebula are present, and resolved. 

In section \ref{sec:obs} we present the ILT observations and describe the ancillary data used in this work. In section \ref{sec:results} we present spectral index maps of the Crab Nebula, and the results for fitting the images at different frequencies as a synchrotron source subject to free-free absorption. We also quote a surface brightness upper limit at LOFAR frequencies for the, as of yet undetected, Crab Nebula supernova remnant shell. In section \ref{sec:discussion} we discuss the implications of the LOFAR observations for the conditions in the absorbing Rayleigh-Taylor filaments, and in section \ref{sec:summary} we summarise our work.

\section{Observations and data reduction}
\label{sec:obs}

\begin{figure*}
    \centering
    \includegraphics[width=0.85\textwidth]{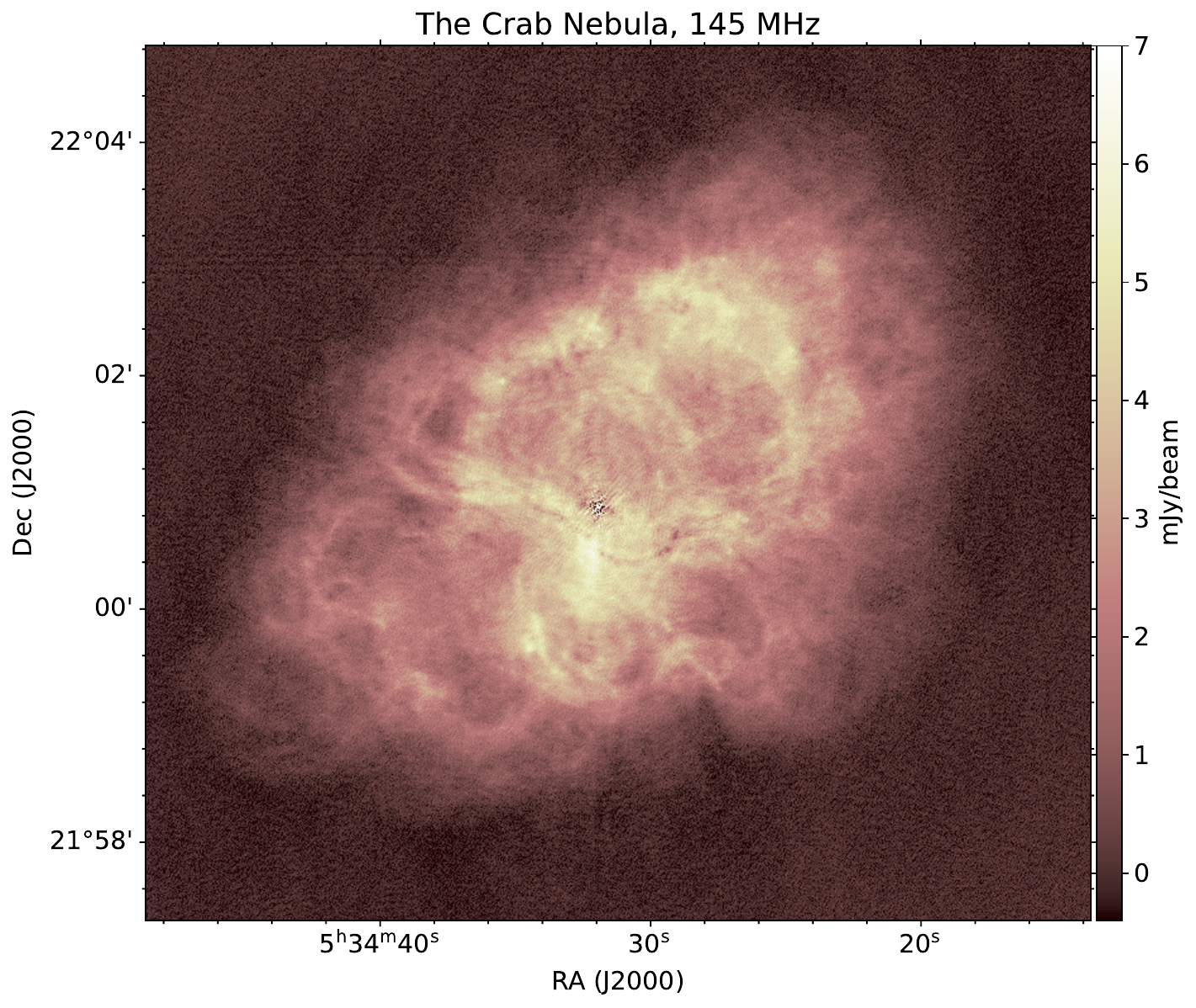}
    \caption{The Crab Nebula as seen with the High Band Antenna (HBA) of the International LOFAR Telescope (ILT). This map has a central frequency of 145~MHz, an angular resolution of $0.43'' \times 0.28''$, an rms noise of 260~$\mu$Jy~bm$^{-1}$, and a dynamic range of 61,000; the area displayed here has a size of $8'\times8'$.}
    \label{fig:crabHBA}
\end{figure*}

\subsection{ILT observations}

The observations used in this work were taken as part of the LOFAR Two-Metre Sky Survey \cite[LoTSS,][]{shimwell17,shimwell19,shimwell22}, a survey with the LOFAR HBA at a central frequency of 145~MHz that covers 85\% of the Northern sky. Although the LoTSS data products are imaged using only the Dutch LOFAR stations, resulting in $6''$ resolution and typically reaching sensitivities of $\sim100~\mu$Jy~bm$^{-1}$, most of the survey data were recorded using the full ILT array. This provides the option of re-processing the data to include the international stations, increasing the maximum baseline length and reaching an angular resolution of $0.3''$.

The LoTSS observation covering the Crab Nebula was taken on the 4th of March 2023 for a total on-source integration time of 4~hours. The data were recorded in full linear polarization mode between 120 and 168~MHz with a frequency channel width of 12~kHz and an integration interval of 1~second. In addition, the primary calibrator source 3C\,48 was observed with the same observing setup for 10~minutes immediately afterwards.

The data on 3C48 were first processed using the \textsc{prefactor} package \citep[now the LOFAR Initial Calibration pipeline, LINC;][]{vanweeren16, williams16, gasperin19} to obtain calibration solutions for the bandpass, the polarization alignment, and the clock offsets for each LOFAR station. Next, these solutions were applied to the target data and an initial phase calibration cycle was performed by comparing the data for the Dutch stations against a sky model obtained from the TIFR GMRT Sky Survey \citep[TGSS,][]{intema17}.

Following this initial calibration, the full international LOFAR array was calibrated using a bespoke strategy implemented using LOFAR-VLBI pipeline tools \citep{morabito21,vanweeren21}. First, the data were phase-shifted to the coordinates of the Crab pulsar, converted to circular polarization, and averaged to a channel width of 49~kHz and an integration interval of 4~seconds. Next, calibration solutions were derived for the full array on the central pulsar using only baselines exceeding 80~k\(\mathrm{\lambda}\) to suppress the surrounding large-scale emission. This baseline cut causes the calibration to only consider emission on scales smaller than \(\sim\)2.6 arcseconds. These calibration solutions include short-timescale phase corrections for left- and right-handed circular polarizations separately, as well as gain corrections to solve for the slowly-varying gain of the array and to correct for giant pulses \citep{heiles70} from the central pulsar. The giant pulses are intense bursts of radio emission occurring on nanosecond time scales, but emitting several orders of magnitude above the typical pulse flux \citep{karuppusamy10}. After deriving all calibration solutions and applying these to the data, the full Crab Nebula was imaged using \textsc{WSClean} \citep{offringa14}. The resulting image is shown in Fig.~\ref{fig:crabHBA}.

\begin{table*}[]
\centering
\begin{tabular}{c|ccccc}
\textbf{Instrument} &
  \textbf{\begin{tabular}[c]{@{}c@{}}Central frequency\\ (MHz)\end{tabular}} &
  \textbf{\begin{tabular}[c]{@{}c@{}}Angular resolution\\ (arcsec$^2$)\end{tabular}} &
  \textbf{\begin{tabular}[c]{@{}c@{}}Flux density \\ (Jy)\end{tabular}} &
  \textbf{Obs. date} &
  \textbf{Reference} \\ 
  \hline
LOFAR LBA  & 54   & $11 \times 8$      & $2218\pm224$ & 3 March 2016 & \cite{degasperin20}             \\
LOFAR HBA  & 145  & $0.43 \times 0.28$ & $1412\pm141$ & 4 March 2023 & This work                                                               \\
VLA L-band & 1520 & 0.84 $\times$ 0.79 & $806\pm80$  & 31 May 2020 & VLA archive \\
& & & & 19 September 2021 & \\
& & & & 20 September 2021 & \\
& & & & 21 March 2022 & \\
& & & & 9 April 2022 & \\
VLA C-band & 5450 & 1.01 $\times$ 0.99 & $545\pm55$  & 20 August 2012 & \cite{bietenholz15}\\
& & & & 26 August 2012 & 
\end{tabular}
\caption{Properties of the images used in this work. The "Flux density" column refers to the integrated flux density of the Crab Nebula, including the pulsar.}
\label{tab:obs}
\end{table*}                                          

\subsection{Ancillary data}

In addition to the LOFAR HBA data described above, we used archival and published radio data in order to conduct a spectral analysis of the source. We used the LOFAR Low-Band Antenna (LBA) observations published in \cite{degasperin20}, centered at 54~MHz. We also used C-band data centered at 5450~MHz taken with the VLA and published in \cite{bietenholz15}. Finally, we downloaded calibrated L-band visibilities from the VLA archive \citep{kent18} and made an image centered at 1520~MHz. The properties of each image are summarised in Table \ref{tab:obs}.

\section{Results}
\label{sec:results}

\begin{figure*}
    \centering
    \includegraphics[width=0.45\textwidth]{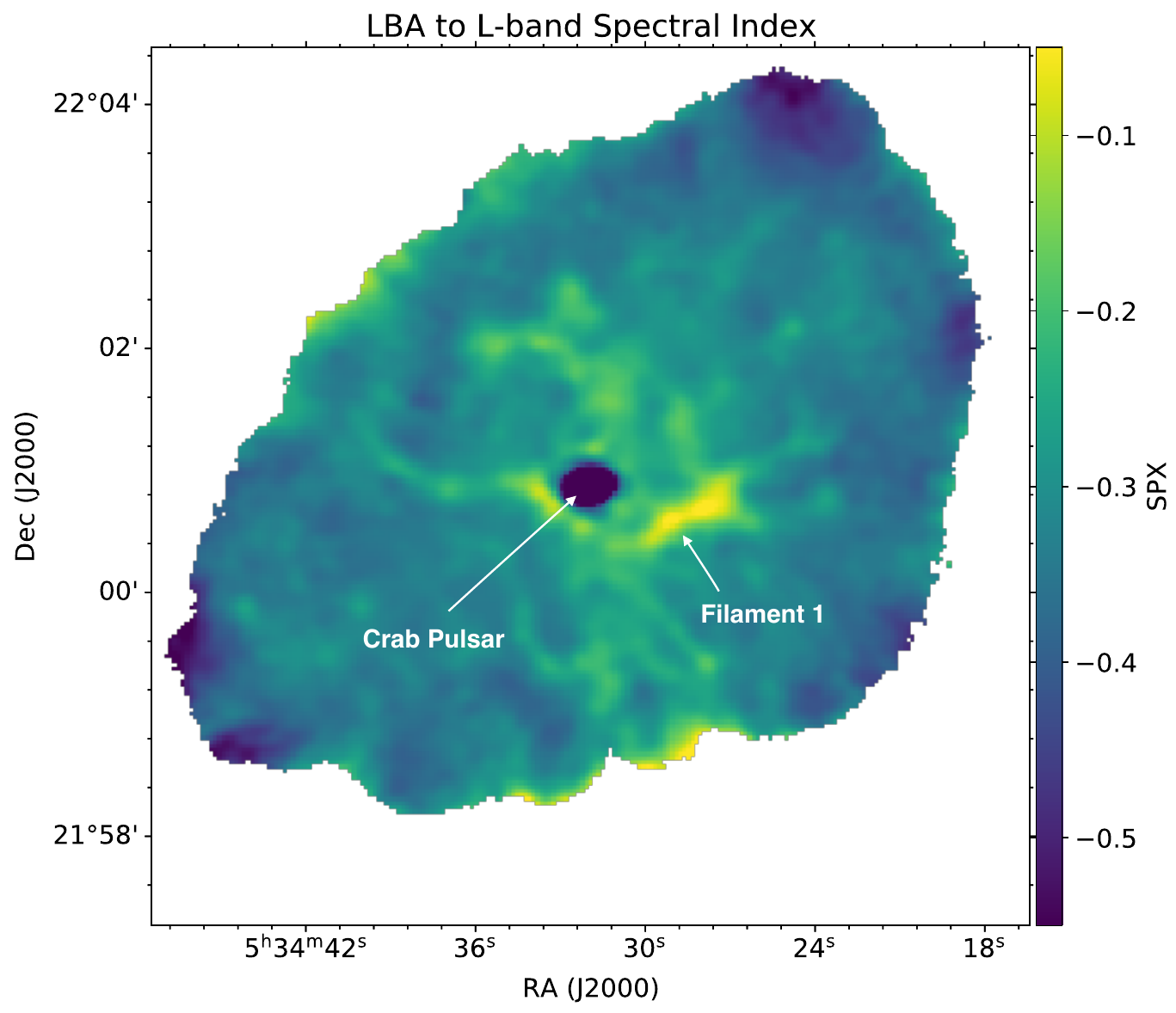}
    \includegraphics[width=0.45\textwidth]{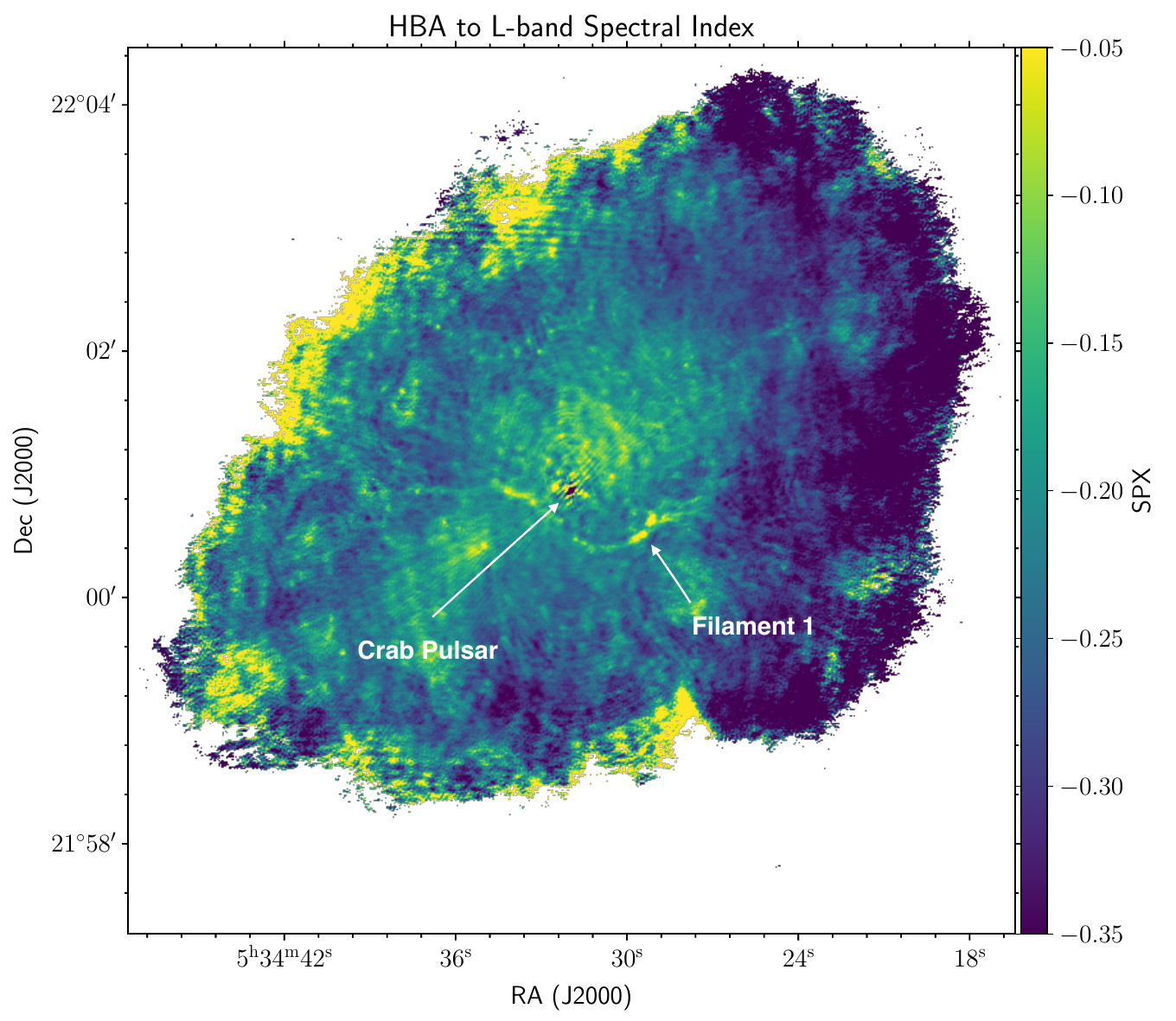}
    \caption{Spectral index maps of the Crab Nebula at low radio frequencies. \textit{Left:} spectral index map between 54~MHz and 1520~MHz, with an angular resolution of $11''$. \textit{Right:} spectral index map between 145~MHz and 1520~MHz, with an angular resolution of $0.8''$.}
    \label{fig:spx}
\end{figure*}

\subsection{Spectral index maps}

To correct for the expansion of the pulsar wind nebula in the time period between the LOFAR and the VLA observations, we spatially rescaled the LOFAR map with respect to the central pulsar. By eye, we estimate a scaling factor of \(0.990 \pm 0.001\) in order to match features in the VLA maps to the LOFAR map better than the angular resolution. This correction agrees well with the difference in age of the Crab Nebula between the VLA and LOFAR epochs \citep{bietenholz15}. With this re-scaling applied, we created spectral index maps between the 145 and 1520~MHz maps (the LOFAR HBA and the VLA L-band), and between the 54 and 1520~MHz maps (LOFAR LBA and the VLA L-band). The results are shown in Fig. \ref{fig:spx}. For each spectral index map, the higher-resolution map was convolved with a Gaussian to match the resolution of the lower-resolution map as given in table \ref{tab:obs}, and then regridded to match its grid. 

\begin{figure*}
    \centering
    \includegraphics[width=0.95\textwidth]{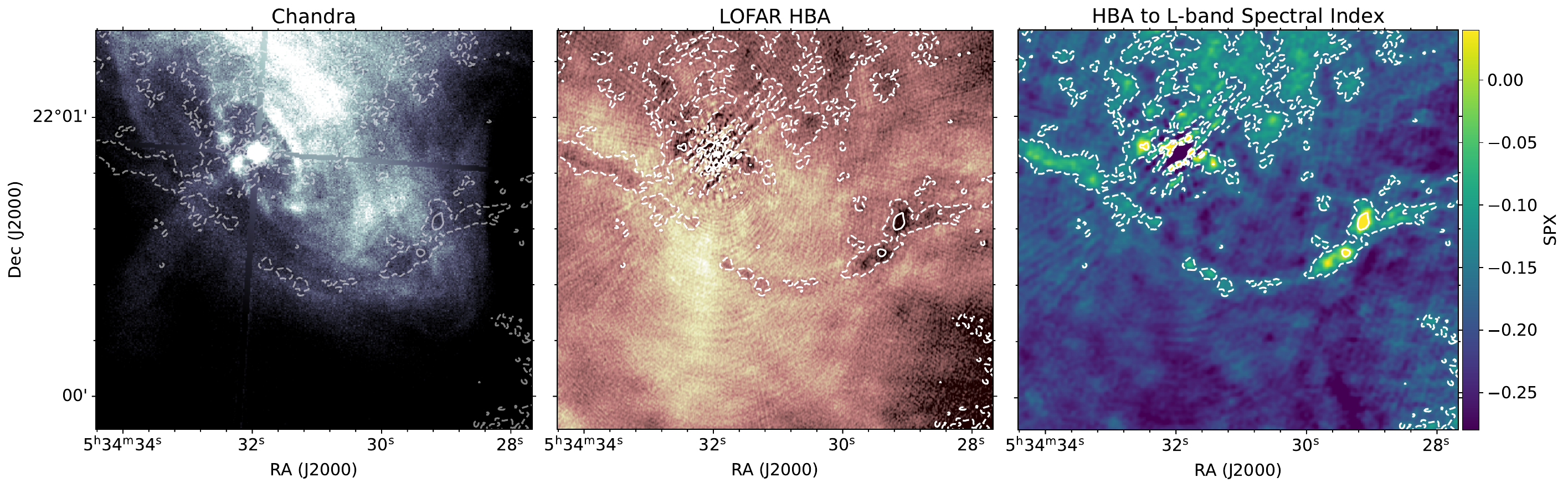}
    \caption{Cutout of the filament with a flat spectral index visible in the HBA to L-band (145~MHz to 1.5~GHz) spectral index map (\lq Filament 1\rq). On the left is the region as seen in the \textit{Chandra} Low-Energy Transmission Grating with the High-Resolution Camera Spectroscopy detector \citep{weisskopf12}, in the middle as seen in the LOFAR HBA at 145~MHz, and in the right, the HBA to L-band (145~MHz to 1.5~GHz) spectral index map (same as Fig. \ref{fig:spx}, right). The contours correspond to the spectral index values of $\alpha=-0.15$ and 0.05.}
    \label{fig:zoom}
\end{figure*}

The most prominent feature in the LBA to L-band (54~MHz to 1.5~GHz) spectral index map is a filament south-west of the pulsar, which stands out as having a flatter spectral index ($\alpha \approx 0$) than the rest of the source (labeled \lq Filament 1\rq\ in Fig. \ref{fig:spx}, left). As mentioned earlier, the presence of an (unresolved) flatter structure at this location was first noted by \cite{bietenholz97} from the analysis of VLA 74~MHz data. The authors attributed the spectral index flattening to free-free absorption by the thermal material in the optical filaments. This interpretation is further confirmed by the morphology of the 54~MHz to 1.5~GHz spectral index map: the regions of flat spectral index in yellow and green clearly trace the outline of the blue-shifted (i.e., foreground) Rayleigh-Taylor fingers as seen in the optical and infrared \cite[see e.g.][]{hester96,loll13,martin21,temim24}. 

Filament 1 is also seen as the main feature in the HBA to L-band (145~MHz to 1.5~GHz) spectral index map: a thin band of emission that partially encircles the pulsar from the south. Figure \ref{fig:zoom} shows a cutout of this feature in the spectral index map, alongside the HBA image and a \textit{Chandra} X-ray map of the source \citep{weisskopf12}. The absorption effects here are so significant that the filament appears as a dark track in the LOFAR HBA total intensity map (Fig. \ref{fig:zoom}, middle) and in the X-rays (Fig. \ref{fig:zoom}, left). Figure \ref{fig:zoom} shows that Filament 1 is in fact inhomogeneous, and has knots of positive spectral index on its Western side.

The HBA to L-band (145~MHz to 1.5~GHz) spectral index map shows a gradient in spectral index seen from the south-east to the north-west. We cannot exclude the possibility that this feature is caused by residual primary beam effects: the source lies one degree from the phase center of the LoTSS pointing that was re-processed to include the international stations and make the map in Fig. \ref{fig:crabHBA}, and hence it is subject to edge effects when applying the primary beam correction. The extended feature of flatter spectral index that follows the semimajor axis of the nebula (coinciding with what is seen as a jet in the X-rays) does not show in the LBA to L-band map (54~MHz to 1.5~GHz) and is also likely due to artifacts in the 145~MHz high-resolution map. 

\subsection{Free-free absorption of the synchrotron nebula by the optical filaments}

The cage of optically-emitting filaments that confines the pulsar wind nebula can cause the latter's synchrotron emission to be free-free absorbed. If the synchrotron emission of the nebula is modulated by free-free absorption, the observed radio emission should have the following functional form \citep{rybicki86}:
\begin{equation}
    S_\nu = S_0 \left(\frac{\nu}{\nu_0}\right)^\alpha e^{-\tau_\nu},
\label{eq:ff}
\end{equation}
with 
\begin{equation}
    \tau_\nu = 3.014 \times 10^4~Z~\left(\frac{T}{\mathrm{K}} \right)^{-3/2} \left(\frac{\nu}{\mathrm{MHz}}\right)^{-2} \left(\frac{EM}{\mathrm{pc~cm}^{-6}}\right) ~g_\mathrm{ff}.
\label{eq:taunu}
\end{equation}
Here, $S_0$ is the flux density of the nebula at some reference frequency $\nu_0$, and $\alpha$ is the value of the spectral index. The free-free optical depth $\tau_\nu$ depends on $Z$ the number of free electrons per absorbing ion (such that $Ze$ is the charge of each ion), $T$ the temperature of the absorbing plasma, $EM$ its emission measure, and $g_\mathrm{ff}$ a Gaunt factor. The emission measure $EM$ is further related to the electron number density $n_e$ via:
\begin{equation}
    n_e = \sqrt{\frac{EM}{l}},
\label{eq:n_e}
\end{equation}
where $l$ is the length of the absorbing slab (in this case, the thickness of the absorbing filament). 

A function of this form behaves like an (unabsorbed) synchrotron power-law at high frequencies, and is optically thick at low frequencies, where the absorption effects are high. Sources with this spectral behaviour show a turnover frequency $\nu_\mathrm{turnover}$ (the value of $\nu$ for which $S_\nu$ peaks) that depends on $T$, $Z$, and $EM$\footnote{The dependency, for constant $T$ and $Z$, is found by setting the derivative of equation \ref{eq:ff} to zero and solving numerically for $\nu$. Note that $\nu_\mathrm{turnover} \approx \nu_{1}$, the value of $\nu$ for which the optical depth is equal to one ($\tau_\nu1=1$).}.

Given the frequency dependence of the optical depth, the effect of absorption is seen most clearly in the map centered at 54~MHz (this image can be seen in Figure 1 of \cite{degasperin20}; the Crab Nebula is labeled as Taurus A). However, note that the filament shown in Fig. \ref{fig:zoom} (Filament 1) shows in some knots a positive spectral index value in the HBA to L-band (145~MHz to 1.5~GHz) map, indicating that the photoionized material is so dense that the turnover happens at frequencies higher than 150~MHz. 

If we assume that the value of the spectral index is fixed at $\alpha=-0.27$ \cite[as suggested by e.g.][]{bietenholz97}, and that the Gaunt factor $g_\mathrm{ff}$ is weakly dependent on $\nu$, we can rewrite equations \ref{eq:ff} and \ref{eq:taunu} as functions of only two variables, the flux density $S$ at a given frequency, e.g. 1~GHz, and a constant $X$ that determines the absorption for a given frequency:
\begin{equation}
    S_\nu = S_{\mathrm{1~GHz}} \left(\frac{\nu}{\mathrm{1~GHz}}\right)^{-0.27} e^{-X\nu^2}
\label{eq:simp}
\end{equation}

\begin{figure*}
    \centering
    \includegraphics[width=0.49\textwidth]{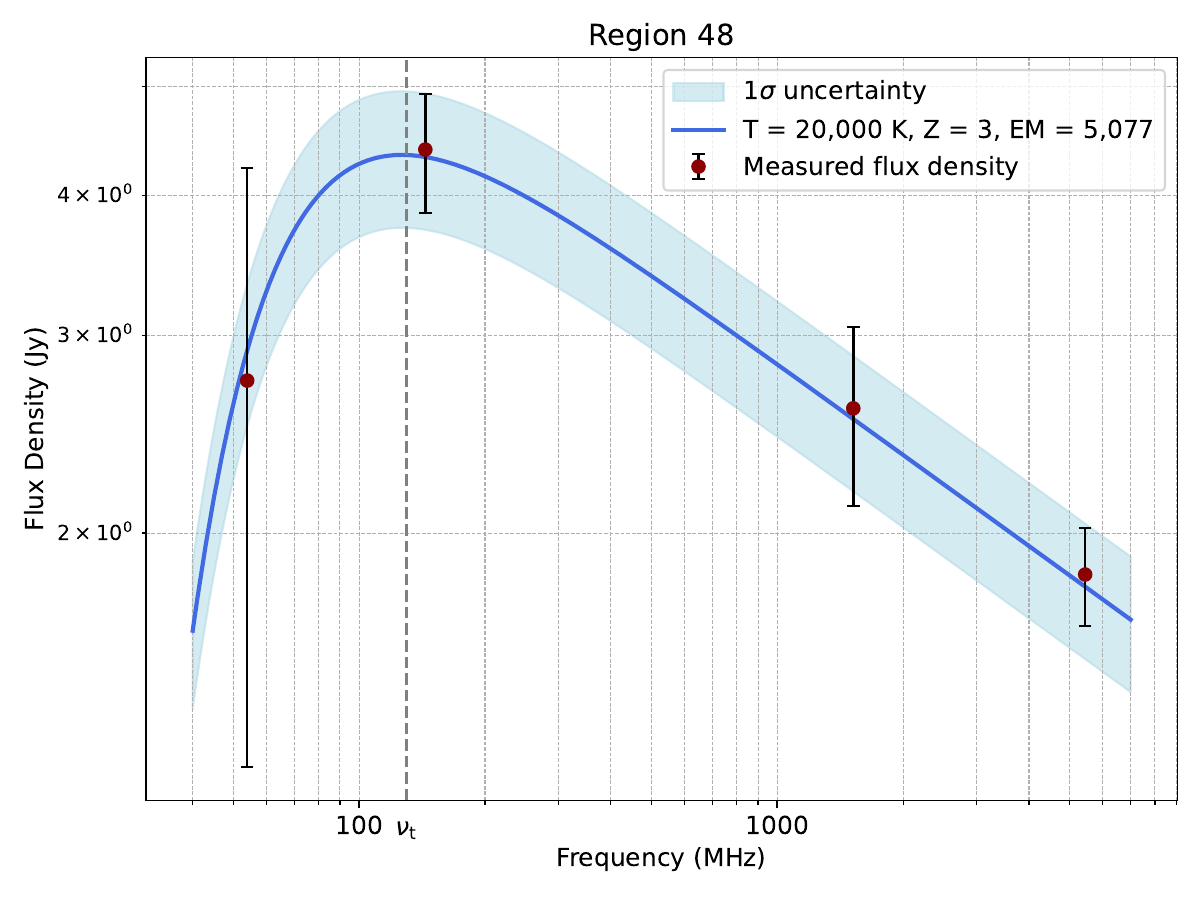}
    \includegraphics[width=0.49\textwidth]{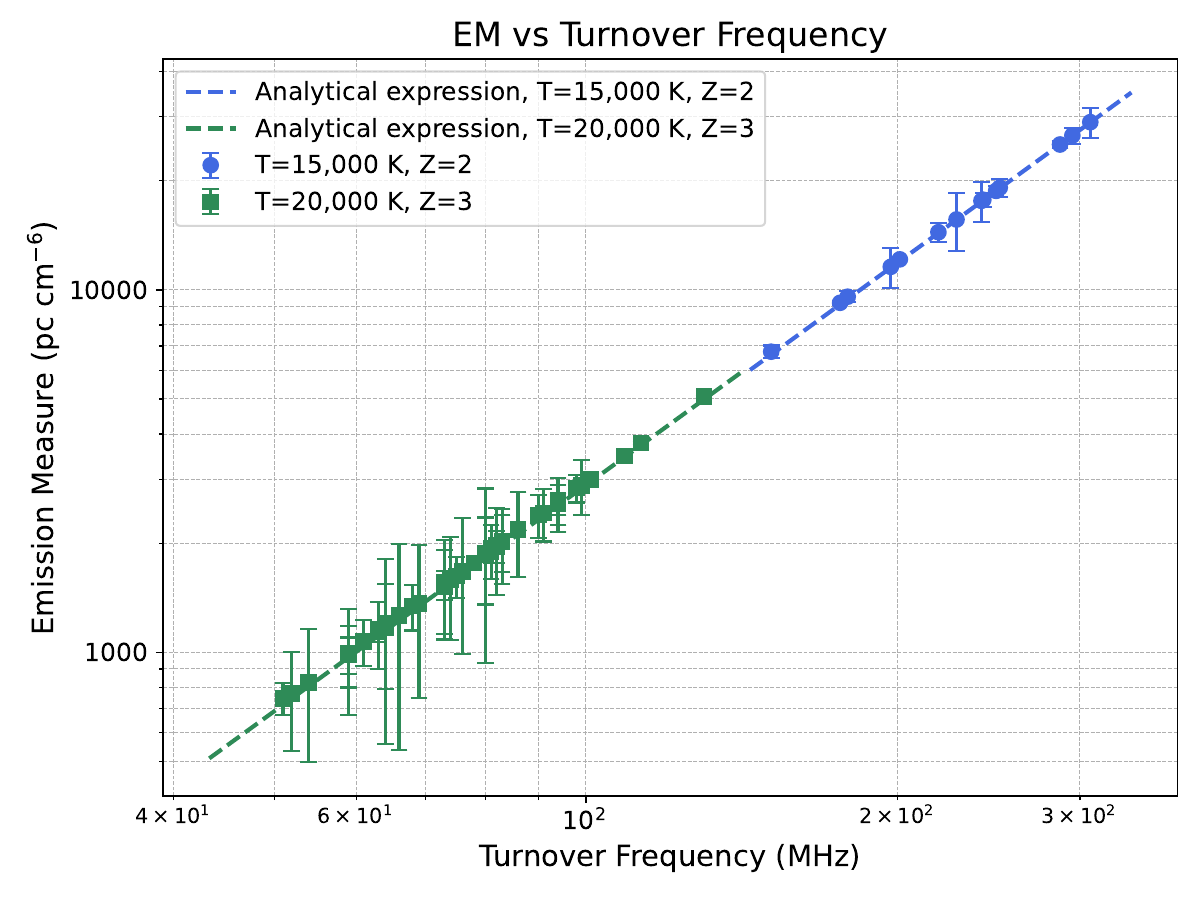}
    \caption{
    \textit{Left:} example of the best-fit procedure for a given region (Region 48 as labeled in Fig. \ref{fig:BF}). The data points are the integrated flux density measurements for the region in each of the images. We fit for the values of $S_0$ and $X$ in equation \ref{eq:simp}, and then convert to a best-fit value of $EM$ with its associated error after fixing, in this case, $T=20,000$~K and $Z=3$. This is the same procedure that is followed on a pixel-by-pixel basis when making Figs. \ref{fig:EM} and \ref{fig:EMLR}. The turnover frequency $\nu_\mathrm{t}$ in this case occurs at 130~MHz, and is indicated by the vertical dashed line.
    \textit{Right:} emission measure $EM$ as a function of turnover frequency $\nu_\mathrm{turnover}$ for the regions labeled in Fig. \ref{fig:BF}. The values in green correspond to the regions in table \ref{tab:LR}, and the values in blue to those in table \ref{tab:HR}. The dashed lines are the analytical relationship between $EM$ and $\nu_\mathrm{turnover}$ for the corresponding values of $T$ and $Z$. }
    \label{fig:nu_turn}
\end{figure*}

\begin{figure*}
    \centering
    \includegraphics[width=\textwidth]{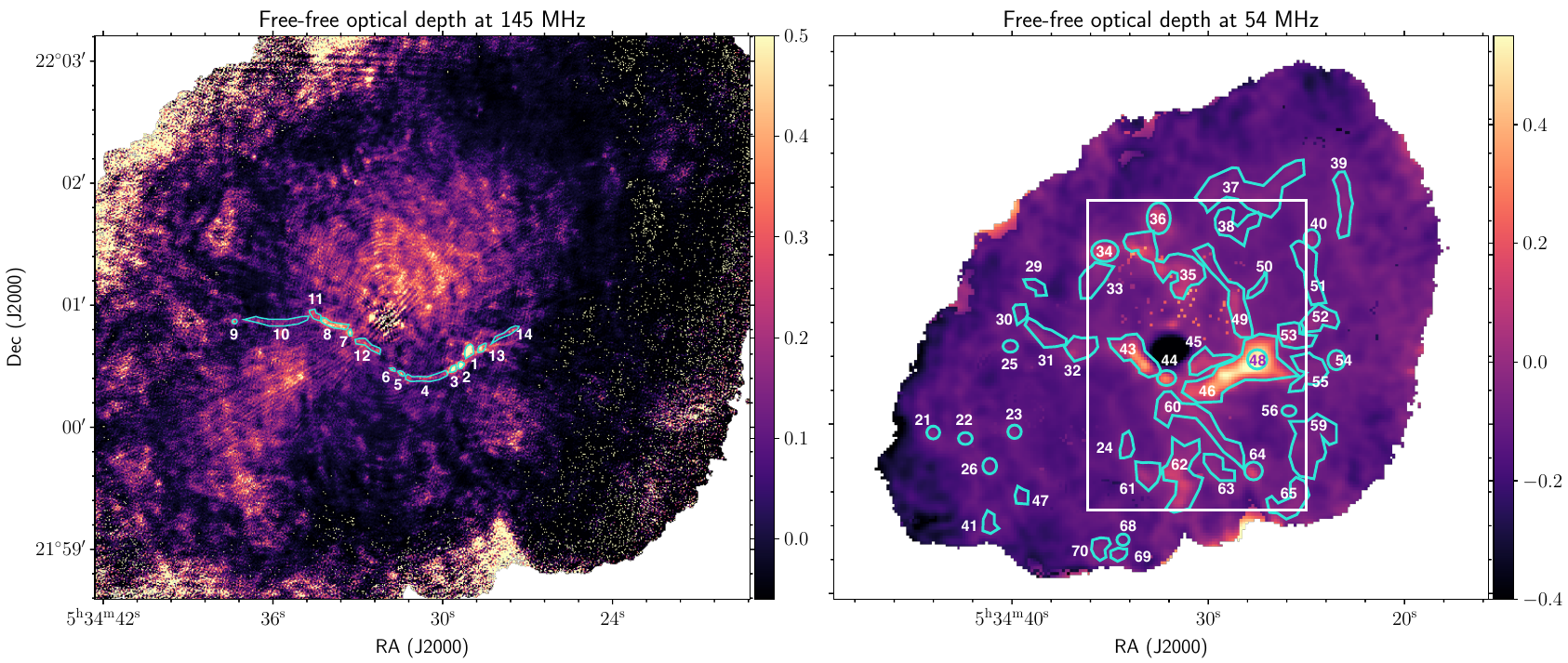}
    \caption{
    \textit{Left:} best-fit free-free optical depth at 145~MHz. The integrated values for the regions labeled $1-14$ are shown in Table \ref{tab:HR}. \textit{Right:} best-fit free-free optical depth at 54~MHz. The integrated values for the regions labeled $21-70$ are shown in table \ref{tab:LR}. The region on the white box corresponds to the inset plotted in Fig. \ref{fig:EMLR}. }
    \label{fig:BF}
\end{figure*}

Since we have images at multiple frequencies (54, 145, 1520, and 5450 MHz), we can fit, for each pixel, for $S_\mathrm{1~GHz}$ and $X$ in equation \ref{eq:simp} to understand the degree of free-free absorption done by the Rayleigh-Taylor fingers (after convolving and regridding the maps to a common resolution, that of the 54~MHz map). An example of the fitting procedure for a region showing high absorption is show in Fig. \ref{fig:nu_turn}, left. In Fig. \ref{fig:BF}, right, we show the 
value of the free-free optical depth at 54~MHz, $\tau_\mathrm{54~MHz}$ for the corresponding best-fit value of $X$. 
Note that for some pixels the value of the optical depth is as high as $\tau_\mathrm{54~MHz} = 0.6 \pm 0.1$.

We can follow a similar approach with the three higher frequency maps to highlight the effects of absorption at higher resolution (in this case, the images are convolved to the resolution of the 1520~MHz map, 1$\arcsec$). The best-fit optical depth at 145~MHz is presented in Fig. \ref{fig:BF}, left. For the heavily absorbed Filament 1, the optical depth values in the high-resolution fit show $\tau_\mathrm{145~MHz}>1$, with values as high as $\tau_\mathrm{145~MHz} = 1.5 \pm 0.1$.

\subsection{Electron density in the Rayleigh-Taylor filaments}
\label{subsec:edens}

Ideally, we would use the best-fit values of Fig. \ref{fig:BF} alongside equations \ref{eq:taunu} and \ref{eq:n_e} to obtain a value for the emission measure $EM$ and electron number density $n_e$ in the absorbing filaments. This requires knowing the average temperature $T$ and charge $Z$ of the absorbing material. Note that neither the neutral atoms, the molecules, nor the dust present in the filaments \citep{graham90,sankrit98,temim24} contribute to the low frequency absorption that this method is sensitive to (since free-free absorption requires free electrons).

A further necessary parameter is the thickness of the absorbing volume. Filament 1 appears to have a clumpy structure (see Fig. \ref{fig:zoom}) , but we measure its average width to be 3\arcsec. In the low-resolution map, we are sensitive to thicker (yet in some cases resolved) structures; for the same Filament 1 we measure an average width of 15\arcsec. Here we will approximate the length of the absorbing slab $l$ in equation \ref{eq:n_e} as the measured thickness of the filaments.

\begin{figure*}
    \centering
    \includegraphics[width=0.65\textwidth]{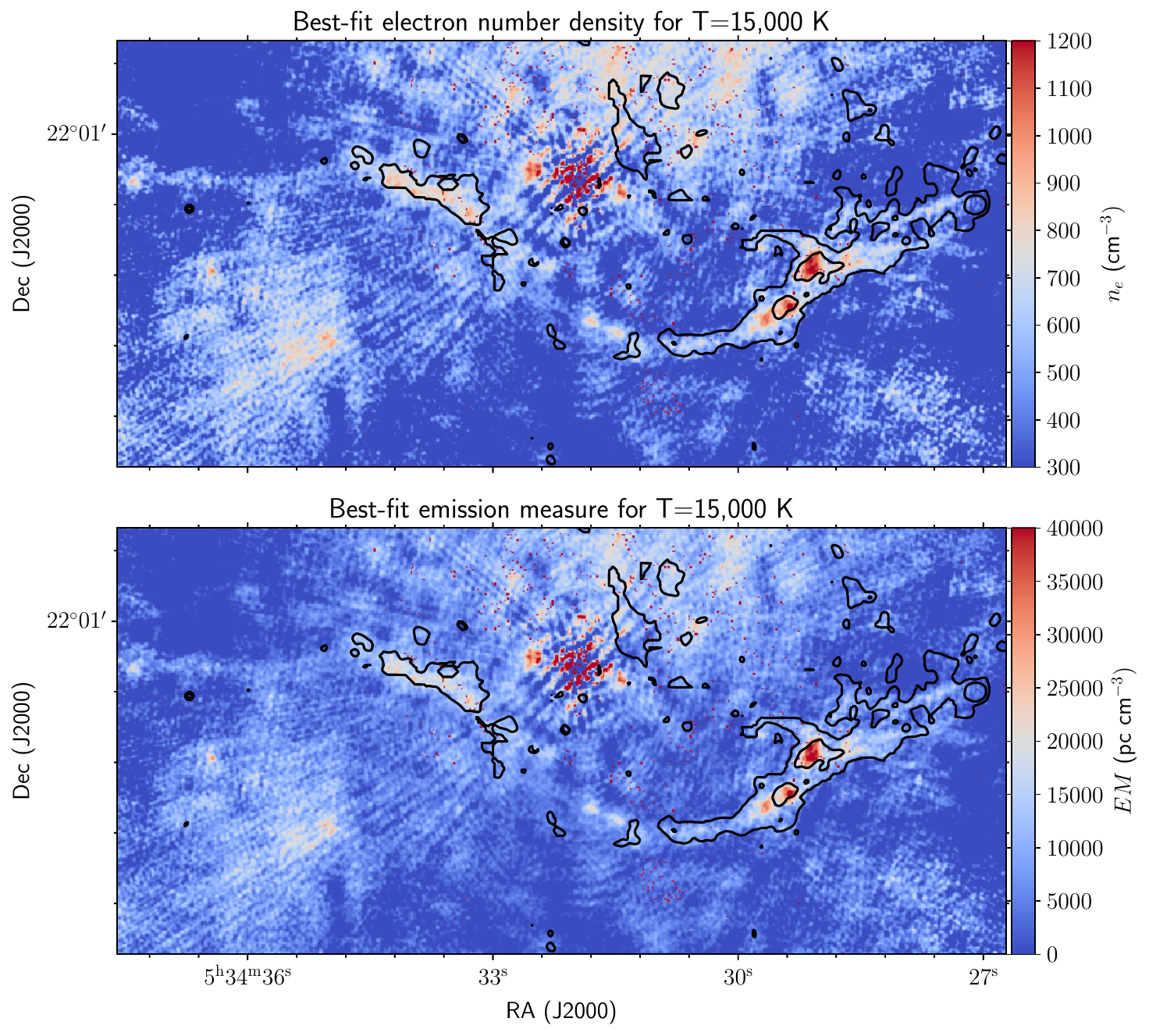}
    \caption{Results from combining our best-fit absorption values with equations \ref{eq:taunu} and \ref{eq:n_e} using $T=15,000$~K \citep{fesen98}, $Z=2$, and $l=0.03$~pc. Plotted are the electron number density $n_e$ (above) and emission measure $EM$ (below). The contours correspond to the blue-shifted oxygen emission as mapped in  \cite{martin21}. Note that the contour levels are the same in Fig. \ref{fig:EMLR}.}
    \label{fig:EM}
\end{figure*}

\cite{fesen98} present spectroscopic observations of several filaments in the Crab Nebula, and find electron temperatures in the range of $11,000-18,300$~K and a typical electron density in the filaments of 1,300~cm$^{-3}$. Using $Z=2$ and $T=15,000$~K and a filament thickness of $l=3\arcsec$ \cite[0.03~pc for a 2~kpc distance to the Crab,][]{trimble73} the resulting  $n_e$
values are in good agreement with \cite{fesen98}: the densest knot has $n_e=1410\pm140$~cm$^{-3}$ at its highest value ($\alpha=5\mathrm{h}34\mathrm{m}29\fs1$, $\delta=+22\degr00\arcmin36.6\arcsec$) and the filaments show $n_e=600-800$~cm$^{-3}$ in their more diffuse parts, see Fig. \ref{fig:EM}. 

We also measured the flux density in a series of regions (shown in Fig. \ref{fig:BF}, left) in each of the images and jointly fitted for the absorption  for their average values of $EM$ and $n_e$. In table \ref{tab:HR} we show the best-fit values for $EM_{Z=2,T=15,000}$ and $n_{e\,\, l=0.03})$, where:
\begin{equation}
    EM(Z,T) = EM_{Z=2,T=15,000} \left(\frac{3}{Z}\right)\left(\frac{T}{20,000~\mathrm{K}}\right)^{3/2}
\end{equation}
and 
\begin{equation}
    n_e(Z, T, l) = n_{e\,\, l=0.03} \left(\frac{2}{Z}\right)^{1/2}\left(\frac{T}{15,000~\mathrm{K}}\right)^{3/4}\left(\frac{0.03~\mathrm{pc}}{l}\right)^{1/2}
\end{equation}

Carrying out a similar analysis using the low-resolution best-fit maps (which include the LOFAR LBA data at 54~MHz) the effects of absorption are visible at scales that are partially resolved in some cases, indicating a larger absorbing volume. As seen in Fig. \ref{fig:BF}, right, the region of most intense absorption is thicker than the angular resolution of the LBA map, with an approximate width of $15\arcsec$ and measuring over $20\arcsec$ at its thickest point. However, many of the features are only as wide as the angular resolution of the LBA map, implying that their thickness should be less than $0.08$~pc.

\begin{figure*}
    \centering
    \includegraphics[width=0.7\textwidth]{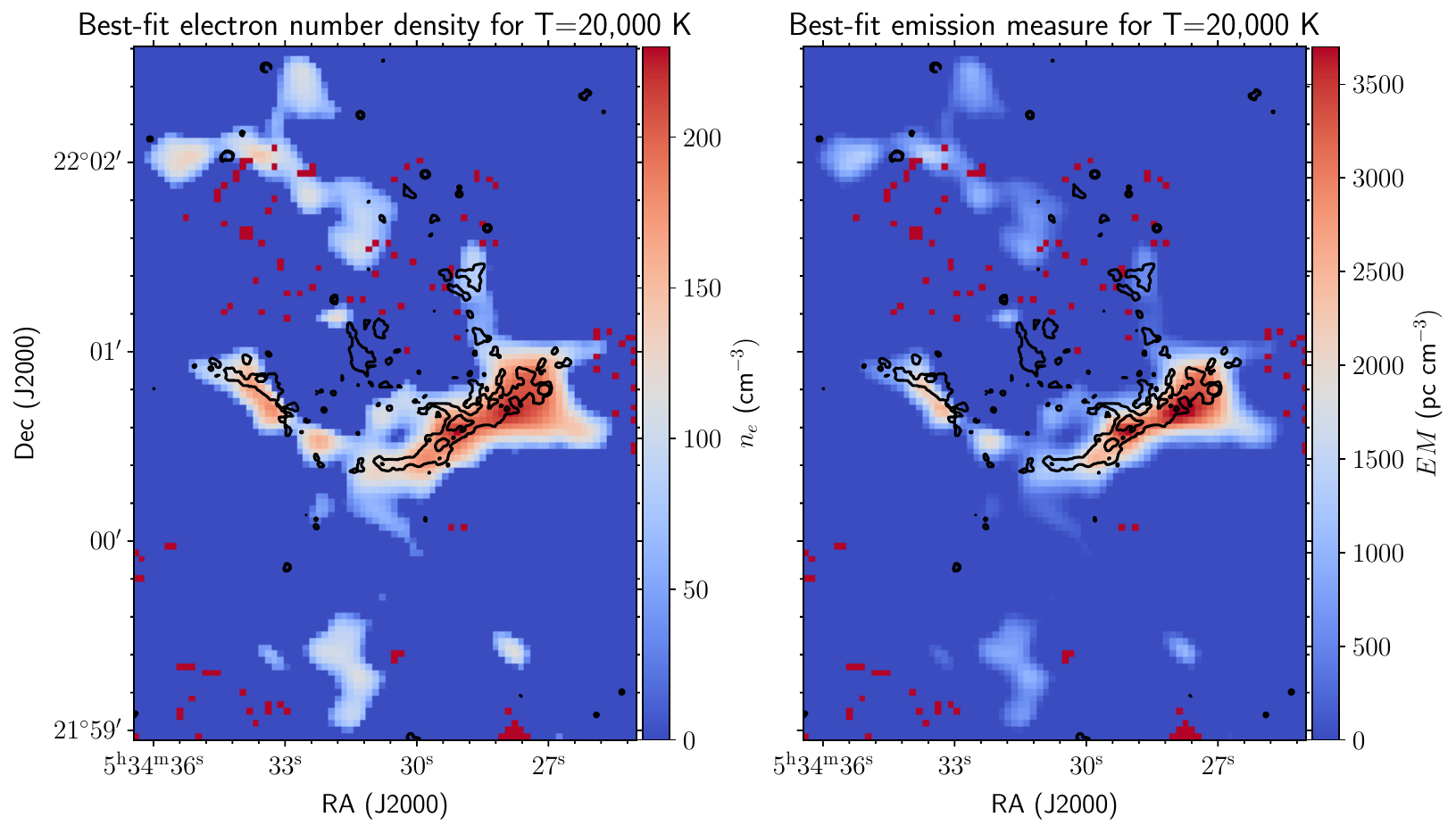}
    \caption{Results from combining our best-fit absorption values with equations \ref{eq:taunu} and \ref{eq:n_e} using $T=20,000$~K \citep{fesen98}, $Z=3$, and $l=0.08$~pc. Plotted are the electron number density $n_e$ (left) and emission measure $EM$ (right). The region displayed here is shown as a blue box in Fig. \ref{fig:BF}, middle. The contours correspond to the blue-shifted oxygen emission as mapped in  \cite{martin21}. Note that the contour levels are the same in Fig. \ref{fig:EM}.}
    \label{fig:EMLR}
\end{figure*}

We do not know the average $T$ and $Z$ values in the volume probed by the LOFAR LBA observations, but from the stratified ionisation structure of the filaments it is clear that both $Z$ and $T$ should be higher than the values we used for the high-resolution analysis. Using $Z=3$ and $T=20,000$ and a filament thickness of $l=8\arcsec$ (0.075~pc) the resulting  $n_e$ are in the range of $\sim50-250$~cm$^{-3}$, see Fig. \ref{fig:EMLR}. For a series of regions shown in Fig. \ref{fig:BF}, right, we show the best-fit values of $EM_{Z=3,T=20,000}$ and $n_{e\,\, l=0.08}$, defined as above, in table \ref{tab:LR}. The flux density values and best-fit parameters for a sample region showing high absorption are plotted in Fig. \ref{fig:nu_turn}, right. Note that for the region 46 (corresponding to Filament 1), which is resolved, the value of $n_e$ is slightly overestimated, as a more appropriate value of $l$ for this region is $l\sim0.15$~pc.

Finally, in tables \ref{tab:HR} and \ref{tab:LR} we show the turnover frequency $\nu_\mathrm{turover}$ corresponding to the peak of the spectrum of each region for the  best-fit values of the parameters. The turnover frequency is obtained directly from fitting equation \ref{eq:simp} to the parameters. Figure \ref{fig:nu_turn}, right, shows a plot of the emission measure as a function of turnover frequency for the regions under consideration.

\begin{table*}[]
\centering
\begin{tabular}{cccccc}
\hline
Region & \begin{tabular}[c]{@{}c@{}}RA \\ (hh:mm:ss)\end{tabular} &
\begin{tabular}[c]{@{}c@{}}Dec \\ (dd:mm:ss)\end{tabular} &
\begin{tabular}[c]{@{}c@{}}$\nu_\mathrm{turnover}$ \\ (MHz)\end{tabular} &
\begin{tabular}[c]{@{}c@{}}$EM_{Z=2,T=15,000}$ \\ (pc~cm$^{-6}$)\end{tabular} &
\begin{tabular}[c]{@{}c@{}} $n_{e\,\, l=0.03}$ \\ (cm$^{-3}$)\end{tabular} \\
\hline
\hline
1 & 5:34:29.0 & +22:00:37.0 & 295 & 26634 $\pm$ 1361 & 956 $\pm$ 161 \\
2 & 5:34:29.4 & +22:00:30.4 & 307 & 28990 $\pm$ 2817 & 998 $\pm$ 173 \\
3 & 5:34:29.6 & +22:00:28.5 & 287 & 25155 $\pm$ 607 & 929 $\pm$ 155 \\
4 & 5:34:30.6 & +22:00:23.7 & 179 & 9569 $\pm$ 339 & 573 $\pm$ 96 \\
5 & 5:34:31.5 & +22:00:26.7 & 219 & 14411 $\pm$ 846 & 703 $\pm$ 119 \\
6 & 5:34:31.8 & +22:00:28.4 & 242 & 17713 $\pm$ 804 & 780 $\pm$ 131 \\
7 & 5:34:33.3 & +22:00:46.2 & 251 & 19123 $\pm$ 1069 & 810 $\pm$ 137 \\
8 & 5:34:33.8 & +22:00:50.5 & 249 & 18737 $\pm$ 577 & 802 $\pm$ 134 \\
9 & 5:34:37.4 & +22:00:51.8 & 228 & 15624 $\pm$ 2845 & 732 $\pm$ 139 \\
10 & 5:34:36.0 & +22:00:52.0 & 151 & 6748 $\pm$ 272 & 481 $\pm$ 80 \\
11 & 5:34:34.5 & +22:00:55.0 & 201 & 12131 $\pm$ 36 & 645 $\pm$ 107 \\
12 & 5:34:32.6 & +22:00:40.0 & 197 & 11568 $\pm$ 1439 & 630 $\pm$ 112 \\
13 & 5:34:28.6 & +22:00:38.9 & 241 & 17590 $\pm$ 2225 & 777 $\pm$ 138 \\
14 & 5:34:29.8 & +22:00:45.6 & 176 & 9205 $\pm$ 124 & 562 $\pm$ 93 \\
\hline
\end{tabular}
\caption{Properties of the regions highlighted in Figure \ref{fig:BF}, right. The column $\nu_\mathrm{turnover}$ corresponds to the best-fit turnover frequency in equation \ref{eq:ff}, where the ionised material becomes optically thick. See section \ref{subsec:edens} for the definitions of $EM_{Z=2,T=15,000}$ and $n_{e\,\, l=0.03}$.}
\label{tab:HR}
\end{table*}

\begin{table*}[]
\centering
\begin{tabular}{cccccc}
\hline
Region & \begin{tabular}[c]{@{}c@{}}RA \\ (hh:mm:ss)\end{tabular} &
\begin{tabular}[c]{@{}c@{}}Dec \\ (dd:mm:ss)\end{tabular} &
\begin{tabular}[c]{@{}c@{}}$\nu_\mathrm{turnover}$ \\ (MHz)\end{tabular} &
\begin{tabular}[c]{@{}c@{}}$EM_{Z=3,T=20,000}$ \\ (pc~cm$^{-6}$)\end{tabular} &
\begin{tabular}[c]{@{}c@{}} $n_{e\,\, l=0.08}$ \\ (cm$^{-3}$)\end{tabular} \\
\hline
\hline
21 & 5:34:44.2 & +21:59:52 & 51 & 747 $\pm$ 76 & 98 $\pm$ 25 \\
22 & 5:34:42.5 & +21:59:48 & 83 & 2016 $\pm$ 470 & 161 $\pm$ 44 \\
23 & 5:34:40.0 & +21:59:53 & 73 & 1535 $\pm$ 141 & 140 $\pm$ 35 \\
24 & 5:34:34.3 & +21:59:42 & 75 & 1623 $\pm$ 212 & 144 $\pm$ 37 \\
25 & 5:34:40.2 & +22:00:55 & 52 & 770 $\pm$ 235 & 99 $\pm$ 29 \\
26 & 5:34:41.2 & +21:59:28 & 68 & 1342 $\pm$ 193 & 131 $\pm$ 34 \\
29 & 5:34:38.8 & +22:01:35 & 63 & 1138 $\pm$ 73 & 121 $\pm$ 30 \\
30 & 5:34:39.7 & +22:01:16 & 59 & 985 $\pm$ 114 & 112 $\pm$ 28 \\
31 & 5:34:38.5 & +22:01:00 & 81 & 1891 $\pm$ 131 & 156 $\pm$ 39 \\
32 & 5:34:36.7 & +22:00:50 & 78 & 1762 $\pm$ 50 & 150 $\pm$ 37 \\
33 & 5:34:35.9 & +22:01:45 & 64 & 1192 $\pm$ 71 & 124 $\pm$ 31 \\
34 & 5:34:35.5 & +22:02:00 & 101 & 2994 $\pm$ 115 & 196 $\pm$ 49 \\
35 & 5:34:32.0 & +22:01:50 & 90 & 2394 $\pm$ 326 & 175 $\pm$ 45 \\
36 & 5:34:32.7 & +22:02:24 & 98 & 2839 $\pm$ 249 & 191 $\pm$ 48 \\
37 & 5:34:27.5 & +22:02:50 & 64 & 1167 $\pm$ 373 & 122 $\pm$ 36 \\
38 & 5:34:28.7 & +22:02:20 & 61 & 1073 $\pm$ 158 & 117 $\pm$ 30 \\
39 & 5:34:23.2 & +22:02:34 & 76 & 1668 $\pm$ 681 & 146 $\pm$ 47 \\
40 & 5:34:24.9 & +22:02:10 & 94 & 2591 $\pm$ 441 & 182 $\pm$ 48 \\
41 & 5:34:41.4 & +21:58:47 & 64 & 1186 $\pm$ 626 & 123 $\pm$ 44 \\
43 & 5:34:34.0 & +22:00:50 & 99 & 2888 $\pm$ 502 & 193 $\pm$ 51 \\
44 & 5:34:32.2 & +22:00:30 & 109 & 3483 $\pm$ 134 & 211 $\pm$ 53 \\
45 & 5:34:30.0 & +22:00:44 & 94 & 2568 $\pm$ 182 & 182 $\pm$ 45 \\
46 & 5:34:30.0 & +22:00:00 & 113 & 3775 $\pm$ 40 & 220 $\pm$ 55 \\
47 & 5:34:39.6 & +21:59:08 & 59 & 990 $\pm$ 190 & 113 $\pm$ 30 \\
48 & 5:34:29.0 & +22:00:33 & 130 & 5077 $\pm$ 44 & 255 $\pm$ 63 \\
49 & 5:34:28.8 & +22:01:29 & 83 & 2025 $\pm$ 361 & 161 $\pm$ 42 \\
50 & 5:34:27.6 & +22:01:34 & 54 & 827 $\pm$ 330 & 103 $\pm$ 33 \\
51 & 5:34:24.6 & +22:01:42 & 59 & 994 $\pm$ 321 & 113 $\pm$ 33 \\
52 & 5:34:24.5 & +22:01:12 & 74 & 1582 $\pm$ 500 & 142 $\pm$ 42 \\
53 & 5:34:26.0 & +22:01:00 & 82 & 1975 $\pm$ 532 & 159 $\pm$ 45 \\
54 & 5:34:23.6 & +22:00:43 & 86 & 2186 $\pm$ 576 & 167 $\pm$ 47 \\
55 & 5:34:24.9 & +22:00:37 & 69 & 1363 $\pm$ 613 & 132 $\pm$ 44 \\
56 & 5:34:26.1 & +22:00:07 & 63 & 1141 $\pm$ 239 & 121 $\pm$ 32 \\
59 & 5:34:25.3 & +21:59:40 & 66 & 1264 $\pm$ 727 & 127 $\pm$ 48 \\
60 & 5:34:30.0 & +22:00:00 & 83 & 2005 $\pm$ 94 & 160 $\pm$ 40 \\
61 & 5:34:33.2 & +21:59:24 & 73 & 1518 $\pm$ 393 & 139 $\pm$ 39 \\
62 & 5:34:31.5 & +21:59:18 & 91 & 2425 $\pm$ 403 & 176 $\pm$ 46 \\
63 & 5:34:29.5 & +21:59:25 & 82 & 1961 $\pm$ 195 & 159 $\pm$ 40 \\
64 & 5:34:27.9 & +21:59:25 & 94 & 2572 $\pm$ 328 & 182 $\pm$ 46 \\
65 & 5:34:26.0 & +21:59:00 & 80 & 1881 $\pm$ 949 & 155 $\pm$ 55 \\
68 & 5:34:35.7 & +21:58:31 & 81 & 1916 $\pm$ 324 & 157 $\pm$ 41 \\
69 & 5:34:34.8 & +21:58:25 & 80 & 1855 $\pm$ 500 & 154 $\pm$ 43 \\
70 & 5:34:35.7 & +21:58:31 & 73 & 1561 $\pm$ 476 & 141 $\pm$ 41 \\
\hline
\end{tabular}
\caption{Properties of the regions highlighted in Figure \ref{fig:BF}, right. The column $\nu_\mathrm{turnover}$ corresponds to the best-fit turnover frequency in equation \ref{eq:ff}, where the ionised material becomes optically thick. See section \ref{subsec:edens} for the definitions of $EM_{Z=3,T=20,000}$ and $n_{e\,\, l=0.08}$.}
\label{tab:LR}
\end{table*}

\subsection{Non-detection of the SNR shell}

The supernova blast wave is not visible as an extended outer shell in the case of the Crab Nebula, unlike for other young SNRs. \cite{sollerman00} found evidence for a fast outer shell surrounding the visible nebula from the far ultraviolet spectrum of the pulsar; they inferred a mass and velocity in the extended shell of $0.3\msun$, and 2,500~km~s$^{-1}$, respectively. 

We made a lower-resolution image in the HBA band with a minimum baseline length of 80$\lambda$, sensitive to structures of up to $43'$, and could not find evidence for a radio shell around the Crab. At a distance of 2~kpc \citep{trimble73}, $43'$ corresponds to a physical length of 25~pc; for reference, if the blast wave had a free expansion velocity of 10,000~km~s$^{-1}$ for 1,000~years, the resulting structure would have a radius of 10~pc ($17'$). 

At the edge of the low-resolution map, $25'$ away from the pulsar, the noise is 4~mJy~bm$^{-1}$ for a $6'' \times 8''$ beam, corresponding to a surface brightness upper limit of $\Sigma = 3.5 \times 10^{-20}$~W~m$^{-2}$~Hz$^{-1}$~sr$^{-1}$; the noise of the map increases as we move closer to the Nebula. This is a less stringent upper limit than the one quoted for the Crab shell by \cite{frail95} from VLA 330~MHz data due to the dynamic range limitations posed by the Crab pulsar, for which we measure a spectral index as steep as $\alpha \sim -3$ between 145 and the 1520~MHz maps.

\section{Discussion}
\label{sec:discussion}

The absorption effects visible in our analysis are associated with features in the optical and infrared that trace the densest material in the filaments. Comparing the dust and [S~\textsc{iii}] emission in Figure 10 of \cite{temim24} with the location of our regions plotted in Fig. \ref{fig:BF} and tables \ref{tab:HR} and \ref{tab:LR}, we can see that all the regions, both in the high and low resolution maps, have a bright counterpart in these components.

It is not straightforward to go from a measurement of the optical depth (which our observations are sensitive to) to physically meaningful quantities describing the plasma. Conclusions about the electron densities in the free-free absorbing material are dependent on its ionisation state and temperature structures. But, as with most astronomical objects, the more high resolution observations are available, the more complex the source appears; in the case of the Crab Nebula, the clearer it becomes that a variety of compositions and ionisation states are present in a given filament, and that the filaments sustain a range of temperatures. This became evident once the \textit{Hubble Space Telescope} (HST) first observed the cage of filaments in the Nebula \citep{sankrit98}, and the complexity of their structure was further revealed by the \textit{James Webb Space Telescope} \cite[JWST,][]{temim24}.
  
\cite{sankrit98} found that the filaments in the Crab show neutral and low-ionisation emission concentrated in sharp structures, and high-ionisation emission occurring in a more diffuse component. This is similar to what we find by comparing the high and low resolution maps: the absorption at 54~MHz is more extended, and overlaps with the majority of the optical foreground (blue-shifted) filaments, whereas the absorption visible in the 145~MHz map is limited to the one filament that traverses the nebula south of the pulsar, Filament 1 (seen in Fig. \ref{fig:EM}). The difference is due to the fact that the high resolution map is only sensitive to structures for which the turnover frequency due to free-free absorption happens at a frequency higher than the LOFAR HBA observations at 145~MHz, which requires overall higher density and/or lower temperature than a turnover at $\gtrsim50$ MHz, which the low-resolution map is sensitive to. 
The regions in high resolution map that we have fitted for have $n_{e\,\, l=0.03}$ in the $500-1,000$~cm$^{-3}$ range (although this is due to integrating over an area; the map in Fig. \ref{fig:EM} shows that for the hotspots the electron density is $>1,000$~cm$^{-3}$), whereas the low-resolution regions have $n_{e\,\, l=0.08} \approx 100-200$~cm$^{-3}$. 

\cite{sankrit98} modeled the photoionisation structure of the filaments with lower temperatures ($4,000-9,000$~K) for the innermost filament; this is less than the value we have assumed to make Fig. \ref{fig:EM} (15,000~K). However, the core of the filaments for which these temperatures are valid has a scale $\sim10^{16}$~cm, whereas for the high-resolution map, our method is only tracking scales of $\sim10^{18}$~cm. 

Our absorption study finds that this density stratification extends to much larger scales than the ones considered by \cite{sankrit98}. Whereas it is the case that many of the regions fitted for in table \ref{tab:LR} have widths of the order of the angular resolution of the LOFAR LBA map, this is not the case for the highly absorbed filament labeled Filament 1, which is partially resolved. Comparing Figs. \ref{fig:EM} and \ref{fig:EMLR} it is clear that the region of high emission measure is more extended in the low-resolution map (it extends further along the western side). Both components are visible within the oxygen contours, but the highest absorption at high frequencies coincides with the brightest optical emission, whereas the highest absorption values in the low-resolution map coincide with a region of more extended and fainter optical emission. 

Whereas the absorbed filament in the high-resolution map is resolved, and has a thickness of $\sim0.03$~pc, in most cases, the filaments visible in the low-resolution map are unresolved in their width dimension, implying that they are in general $<0.08$~pc. However, at its thickest, the region of high optical depth at 54~MHz is $\sim0.2$~pc. It is not possible to tell whether this is due to an overlap of thinner filaments as opposed to an extended envelope, although if this were the case, we would expect to see their structure in the high-resolution map. The lack of features in the optical depth map at 145~MHz points to the $\sim0.2$~pc structure being indeed an extended region of relatively low density.

\section{Summary and conclusions}
\label{sec:summary}

We have made a subarcsecond angular resolution map of the Crab Nebula at 145~MHz by re-processing LoTSS observations to include the LOFAR international stations. These are the highest angular resolution low-frequency observations of the source to date. 

Combining this map with archival observations at 5~GHz, 1.5~GHz, and 54~MHz, we have fitted for free-free absorption due to ionised material in front of the synchrotron nebula. Our absorption fit recovered the morphology of the cage of thermally emitting filaments composed of stellar ejecta that surround the pulsar wind nebula, indicating that they are the source of absorption. 
The inclusion or not of the 54~MHz map in the fit (whose resolution is much lower than the remaining three maps, but which is much more sensitive to the effects of absorption) resulted in two different optical depth maps, sensitive to two different filamentary components: a compact component, resolved, with a thickness of $\sim0.03$~pc and electron densities of $500-1,200$~cm$^{-3}$, and a more extended component enveloping the dense component, with electron densities $<200$~cm$^{-3}$. The width of this second component varies, with structures that could well be $<0.08$~pc and structures as large as $0.2$~pc.

These findings suggest that the well-known stratified density/temperature/ionisation structure of the Crab Nebula's optical filaments might extend to larger scales than previously considered.

\vspace{0.2cm}
\small
\noindent\textit{Data Availability.} The 145 and 1520~MHz maps are
are available at the CDS via anonymous ftp to \url{cdsarc.u-strasbg.fr} (130.79.128.5) or via \url{http://cdsweb.u-strasbg.fr/cgi-bin/qcat?J/A+A/}.

\begin{acknowledgements}
We thank M. Bietenholz for sharing his VLA 5~GHz map of the Crab. We also thank the anonymous referee for their useful comments, which improved the paper. MA acknowledges support from the VENI research programme with project number 202.143, which is financed by the Netherlands Organisation for Scientific Research (NWO). 
RT is grateful for support from the UKRI Future Leaders Fellowship (grant MR/T042842/1). This work was supported by the STFC [grants ST/T000244/1, ST/V002406/1]. FS appreciates the support of STFC [ST/Y004159/1].
LOFAR \citep{vanhaarlem13} is the LOw Frequency ARray designed and constructed by ASTRON. It has observing, data processing, and data storage facilities in several countries, which are owned by various parties (each with their own funding sources), and are collectively operated by the ILT foundation under a joint scientific policy. The ILT resources have benefited from the following recent major funding sources: CNRS-INSU, Observatoire de Paris and Universit\'{e} d'Orl\'{e}ans, France; BMBF, MIWF-NRW, MPG, Germany; Science Foundation Ireland (SFI), Department of Business, Enterprise and Innovation (DBEI), Ireland; NWO, The Netherlands; The Science and Technology Facilities Council, UK; Ministry of Science and Higher Education, Poland; Istituto Nazionale di Astrofisica (INAF), Italy.
This research made use of the Dutch national e-infrastructure with support of the SURF Cooperative (e-infra 180169) and the LOFAR e-infra group. The J\"ulich LOFAR Long Term Archive and the German LOFAR network are both coordinated and operated by the J\"ulich Supercomputing Centre (JSC), and computing resources on the supercomputer JUWELS at JSC were provided by the Gauss Centre for Supercomputing e.V. (grant CHTB00) through the John von Neumann Institute for Computing (NIC). This research made use of the University of Hertfordshire high-performance computing facility and the LOFAR-UK computing facility located at the University of Hertfordshire and supported by STFC [ST/P000096/1], and of the Italian LOFAR IT computing infrastructure supported and operated by INAF, and by the Physics Department of Turin University (under an agreement with Consorzio Interuniversitario per la Fisica Spaziale) at the C3S Supercomputing Centre, Italy. 
 
\end{acknowledgements}

\bibliographystyle{aa}

\begin{thebibliography}{39}
\expandafter\ifx\csname natexlab\endcsname\relax\def\natexlab#1{#1}\fi

\bibitem[{{Arias} {et~al.}(2018){Arias}, {Vink}, {de Gasperin}, {Salas},
  {Oonk}, {van Weeren}, {van Amesfoort}, {Anderson}, {Beck}, {Bell}, {Bentum},
  {Best}, {Blaauw}, {Breitling}, {Broderick}, {Brouw}, {Br{\"u}ggen},
  {Butcher}, {Ciardi}, {de Geus}, {Deller}, {van Dijk}, {Duscha},
  {Eisl{\"o}ffel}, {Garrett}, {Grie{\ss}meier}, {Gunst}, {van Haarlem},
  {Heald}, {Hessels}, {H{\"o}randel}, {Holties}, {van der Horst}, {Iacobelli},
  {Juette}, {Krankowski}, {van Leeuwen}, {Mann}, {McKay-Bukowski}, {McKean},
  {Mulder}, {Nelles}, {Orru}, {Paas}, {Pandey-Pommier}, {Pandey}, {Pekal},
  {Pizzo}, {Polatidis}, {Reich}, {R{\"o}ttgering}, {Rothkaehl}, {Schwarz},
  {Smirnov}, {Soida}, {Steinmetz}, {Tagger}, {Thoudam}, {Toribio}, {Vocks},
  {van der Wiel}, {Wijers}, {Wucknitz}, {Zarka}, \& {Zucca}}]{arias18}
{Arias}, M., {Vink}, J., {de Gasperin}, F., {et~al.} 2018, \aap, 612, A110

\bibitem[{{Arias} {et~al.}(2019){Arias}, {Vink}, {Zhou}, {de Gasperin},
  {Hardcastle}, \& {Shimwell}}]{arias19}
{Arias}, M., {Vink}, J., {Zhou}, P., {et~al.} 2019, \aj, 158, 253

\bibitem[{{Baars} {et~al.}(1977){Baars}, {Genzel}, {Pauliny-Toth}, \&
  {Witzel}}]{baars77}
{Baars}, J.~W.~M., {Genzel}, R., {Pauliny-Toth}, I.~I.~K., \& {Witzel}, A.
  1977, \aap, 61, 99

\bibitem[{{Bietenholz} {et~al.}(2004){Bietenholz}, {Hester}, {Frail}, \&
  {Bartel}}]{bietenholz04}
{Bietenholz}, M.~F., {Hester}, J.~J., {Frail}, D.~A., \& {Bartel}, N. 2004,
  \apj, 615, 794

\bibitem[{{Bietenholz} {et~al.}(1997){Bietenholz}, {Kassim}, {Frail}, {Perley},
  {Erickson}, \& {Hajian}}]{bietenholz97}
{Bietenholz}, M.~F., {Kassim}, N., {Frail}, D.~A., {et~al.} 1997, \apj, 490,
  291

\bibitem[{{Bietenholz} \& {Kronberg}(1990)}]{bietenholz90}
{Bietenholz}, M.~F. \& {Kronberg}, P.~P. 1990, \apjl, 357, L13

\bibitem[{{Bietenholz} {et~al.}(2015){Bietenholz}, {Yuan}, {Buehler},
  {Lobanov}, \& {Blandford}}]{bietenholz15}
{Bietenholz}, M.~F., {Yuan}, Y., {Buehler}, R., {Lobanov}, A.~P., \&
  {Blandford}, R. 2015, \mnras, 446, 205

\bibitem[{{Blair} {et~al.}(1992){Blair}, {Long}, {Vancura}, {Bowers}, {Conger},
  {Davidsen}, {Kriss}, \& {Henry}}]{blair92}
{Blair}, W.~P., {Long}, K.~S., {Vancura}, O., {et~al.} 1992, \apj, 399, 611

\bibitem[{{Chevalier} \& {Gull}(1975)}]{chevalier75}
{Chevalier}, R.~A. \& {Gull}, T.~R. 1975, \apj, 200, 399

\bibitem[{de~Gasperin {et~al.}(2019)de~Gasperin, Dijkema, Drabent,
  {et~al.}}]{gasperin19}
de~Gasperin, F., Dijkema, T.~J., Drabent, A., {et~al.} 2019, A\&A, 622, A5

\bibitem[{{de Gasperin} {et~al.}(2020){de Gasperin}, {Vink}, {McKean},
  {Asgekar}, {Avruch}, {Bentum}, {Blaauw}, {Bonafede}, {Broderick},
  {Br{\"u}ggen}, {Breitling}, {Brouw}, {Butcher}, {Ciardi}, {Cuciti}, {de Vos},
  {Duscha}, {Eisl{\"o}ffel}, {Engels}, {Fallows}, {Franzen}, {Garrett},
  {Gunst}, {H{\"o}randel}, {Heald}, {Hoeft}, {Iacobelli}, {Koopmans},
  {Krankowski}, {Maat}, {Mann}, {Mevius}, {Miley}, {Morganti}, {Nelles},
  {Norden}, {Offringa}, {Orr{\'u}}, {Paas}, {Pandey}, {Pandey-Pommier},
  {Pekal}, {Pizzo}, {Reich}, {Rowlinson}, {Rottgering}, {Schwarz}, {Shulevski},
  {Smirnov}, {Sobey}, {Soida}, {Steinmetz}, {Tagger}, {Toribio}, {van Ardenne},
  {van der Horst}, {van Haarlem}, {van Weeren}, {Vocks}, {Wucknitz}, {Zarka},
  \& {Zucca}}]{degasperin20}
{de Gasperin}, F., {Vink}, J., {McKean}, J.~P., {et~al.} 2020, \aap, 635, A150

\bibitem[{{Fesen} \& {Kirshner}(1982)}]{fesen98}
{Fesen}, R.~A. \& {Kirshner}, R.~P. 1982, \apj, 258, 1

\bibitem[{{Frail} {et~al.}(1995){Frail}, {Kassim}, {Cornwell}, \&
  {Goss}}]{frail95}
{Frail}, D.~A., {Kassim}, N.~E., {Cornwell}, T.~J., \& {Goss}, W.~M. 1995,
  \apjl, 454, L129

\bibitem[{{Graham} {et~al.}(1990){Graham}, {Wright}, \& {Longmore}}]{graham90}
{Graham}, J.~R., {Wright}, G.~S., \& {Longmore}, A.~J. 1990, \apj, 352, 172

\bibitem[{{Heiles} \& {Campbell}(1970)}]{heiles70}
{Heiles}, C. \& {Campbell}, D.~B. 1970, \nat, 226, 529

\bibitem[{{Hester}(2008)}]{hester08}
{Hester}, J.~J. 2008, \araa, 46, 127

\bibitem[{{Hester} {et~al.}(1996){Hester}, {Stone}, {Scowen}, {Jun},
  {Gallagher}, {Norman}, {Ballester}, {Burrows}, {Casertano}, {Clarke},
  {Crisp}, {Griffiths}, {Hoessel}, {Holtzman}, {Krist}, {Mould}, {Sankrit},
  {Stapelfeldt}, {Trauger}, {Watson}, \& {Westphal}}]{hester96}
{Hester}, J.~J., {Stone}, J.~M., {Scowen}, P.~A., {et~al.} 1996, \apj, 456, 225

\bibitem[{Intema {et~al.}(2017)Intema, Jagannathan, Mooley, \&
  Frail}]{intema17}
Intema, H.~T., Jagannathan, P., Mooley, K.~P., \& Frail, D.~A. 2017, A\&A, 598,
  A78

\bibitem[{{Karuppusamy} {et~al.}(2010){Karuppusamy}, {Stappers}, \& {van
  Straten}}]{karuppusamy10}
{Karuppusamy}, R., {Stappers}, B.~W., \& {van Straten}, W. 2010, \aap, 515, A36

\bibitem[{{Kassim} {et~al.}(1993){Kassim}, {Perley}, {Erickson}, \&
  {Dwarakanath}}]{kassim93}
{Kassim}, N.~E., {Perley}, R.~A., {Erickson}, W.~C., \& {Dwarakanath}, K.~S.
  1993, \aj, 106, 2218

\bibitem[{{Kent} {et~al.}(2018){Kent}, {Masters}, {Chandler}, {Davis}, {Kern},
  {Ott}, {Schinzel}, {Medlin}, {Muders}, {Williams}, {Geers}, {Momjian},
  {Butler}, {Nakazato}, \& {Sugimoto}}]{kent18}
{Kent}, B.~R., {Masters}, J.~S., {Chandler}, C.~J., {et~al.} 2018, in American
  Astronomical Society Meeting Abstracts, Vol. 231, American Astronomical
  Society Meeting Abstracts \#231, 342.14

\bibitem[{{Loll} {et~al.}(2013){Loll}, {Desch}, {Scowen}, \& {Foy}}]{loll13}
{Loll}, A.~M., {Desch}, S.~J., {Scowen}, P.~A., \& {Foy}, J.~P. 2013, \apj,
  765, 152

\bibitem[{{Martin} {et~al.}(2021){Martin}, {Milisavljevic}, \&
  {Drissen}}]{martin21}
{Martin}, T., {Milisavljevic}, D., \& {Drissen}, L. 2021, \mnras, 502, 1864

\bibitem[{Morabito {et~al.}(2022)Morabito, Jackson, Mooney,
  {et~al.}}]{morabito21}
Morabito, L., Jackson, N., Mooney, S., {et~al.} 2022, A\&A, 658, A1

\bibitem[{Offringa {et~al.}(2014)Offringa, McKinley, Hurley-Walker,
  {et~al.}}]{offringa14}
Offringa, A.~R., McKinley, B., Hurley-Walker, N., {et~al.} 2014, MNRAS, 444,
  606

\bibitem[{{Owen} \& {Barlow}(2015)}]{owen15}
{Owen}, P.~J. \& {Barlow}, M.~J. 2015, \apj, 801, 141

\bibitem[{{Rybicki} \& {Lightman}(1986)}]{rybicki86}
{Rybicki}, G.~B. \& {Lightman}, A.~P. 1986, {Radiative Processes in
  Astrophysics}

\bibitem[{{Sankrit} {et~al.}(1998){Sankrit}, {Hester}, {Scowen}, {Ballester},
  {Burrows}, {Clarke}, {Crisp}, {Evans}, {Gallagher}, {Griffiths}, {Hoessel},
  {Holtzman}, {Krist}, {Mould}, {Stapelfeldt}, {Trauger}, {Watson}, \&
  {Westphal}}]{sankrit98}
{Sankrit}, R., {Hester}, J.~J., {Scowen}, P.~A., {et~al.} 1998, \apj, 504, 344

\bibitem[{{Shimwell} {et~al.}(2022){Shimwell}, {Hardcastle}, {Tasse}, {Best},
  {R{\"o}ttgering}, {Williams}, {Botteon}, {Drabent}, {Mechev}, {Shulevski},
  {van Weeren}, {Bester}, {Br{\"u}ggen}, {Brunetti}, {Callingham}, {Chy{\.z}y},
  {Conway}, {Dijkema}, {Duncan}, {de Gasperin}, {Hale}, {Haverkorn}, {Hugo},
  {Jackson}, {Mevius}, {Miley}, {Morabito}, {Morganti}, {Offringa}, {Oonk},
  {Rafferty}, {Sabater}, {Smith}, {Schwarz}, {Smirnov}, {O'Sullivan},
  {Vedantham}, {White}, {Albert}, {Alegre}, {Asabere}, {Bacon}, {Bonafede},
  {Bonnassieux}, {Brienza}, {Bilicki}, {Bonato}, {Calistro Rivera}, {Cassano},
  {Cochrane}, {Croston}, {Cuciti}, {Dallacasa}, {Danezi}, {Dettmar}, {Di
  Gennaro}, {Edler}, {En{\ss}lin}, {Emig}, {Franzen}, {Garc{\'\i}a-Vergara},
  {Grange}, {G{\"u}rkan}, {Hajduk}, {Heald}, {Heesen}, {Hoang}, {Hoeft},
  {Horellou}, {Iacobelli}, {Jamrozy}, {Jeli{\'c}}, {Kondapally}, {Kukreti},
  {Kunert-Bajraszewska}, {Magliocchetti}, {Mahatma}, {Ma{\l}ek}, {Mandal},
  {Massaro}, {Meyer-Zhao}, {Mingo}, {Mostert}, {Nair}, {Nakoneczny},
  {Nikiel-Wroczy{\'n}ski}, {Orr{\'u}}, {Pajdosz-{\'S}mierciak}, {Pasini},
  {Prandoni}, {van Piggelen}, {Rajpurohit}, {Retana-Montenegro}, {Riseley},
  {Rowlinson}, {Saxena}, {Schrijvers}, {Sweijen}, {Siewert}, {Timmerman},
  {Vaccari}, {Vink}, {West}, {Wo{\l}owska}, {Zhang}, \& {Zheng}}]{shimwell22}
{Shimwell}, T.~W., {Hardcastle}, M.~J., {Tasse}, C., {et~al.} 2022, \aap, 659,
  A1

\bibitem[{{Shimwell} {et~al.}(2017){Shimwell}, {R{\"o}ttgering}, {Best},
  {Williams}, {Dijkema}, {de Gasperin}, {Hardcastle}, {Heald}, {Hoang},
  {Horneffer}, {Intema}, {Mahony}, {Mandal}, {Mechev}, {Morabito}, {Oonk},
  {Rafferty}, {Retana-Montenegro}, {Sabater}, {Tasse}, {van Weeren},
  {Br{\"u}ggen}, {Brunetti}, {Chy{\.z}y}, {Conway}, {Haverkorn}, {Jackson},
  {Jarvis}, {McKean}, {Miley}, {Morganti}, {White}, {Wise}, {van Bemmel},
  {Beck}, {Brienza}, {Bonafede}, {Calistro Rivera}, {Cassano}, {Clarke},
  {Cseh}, {Deller}, {Drabent}, {van Driel}, {Engels}, {Falcke}, {Ferrari},
  {Fr{\"o}hlich}, {Garrett}, {Harwood}, {Heesen}, {Hoeft}, {Horellou},
  {Israel}, {Kapi{\'n}ska}, {Kunert-Bajraszewska}, {McKay}, {Mohan},
  {Orr{\'u}}, {Pizzo}, {Prandoni}, {Schwarz}, {Shulevski}, {Sipior}, {Smith},
  {Sridhar}, {Steinmetz}, {Stroe}, {Varenius}, {van der Werf}, {Zensus}, \&
  {Zwart}}]{shimwell17}
{Shimwell}, T.~W., {R{\"o}ttgering}, H.~J.~A., {Best}, P.~N., {et~al.} 2017,
  \aap, 598, A104

\bibitem[{{Shimwell} {et~al.}(2019){Shimwell}, {Tasse}, {Hardcastle}, {Mechev},
  {Williams}, {Best}, {R{\"o}ttgering}, {Callingham}, {Dijkema}, {de Gasperin},
  {Hoang}, {Hugo}, {Mirmont}, {Oonk}, {Prandoni}, {Rafferty}, {Sabater},
  {Smirnov}, {van Weeren}, {White}, {Atemkeng}, {Bester}, {Bonnassieux},
  {Br{\"u}ggen}, {Brunetti}, {Chy{\.z}y}, {Cochrane}, {Conway}, {Croston},
  {Danezi}, {Duncan}, {Haverkorn}, {Heald}, {Iacobelli}, {Intema}, {Jackson},
  {Jamrozy}, {Jarvis}, {Lakhoo}, {Mevius}, {Miley}, {Morabito}, {Morganti},
  {Nisbet}, {Orr{\'u}}, {Perkins}, {Pizzo}, {Schrijvers}, {Smith}, {Vermeulen},
  {Wise}, {Alegre}, {Bacon}, {van Bemmel}, {Beswick}, {Bonafede}, {Botteon},
  {Bourke}, {Brienza}, {Calistro Rivera}, {Cassano}, {Clarke}, {Conselice},
  {Dettmar}, {Drabent}, {Dumba}, {Emig}, {En{\ss}lin}, {Ferrari}, {Garrett},
  {G{\'e}nova-Santos}, {Goyal}, {G{\"u}rkan}, {Hale}, {Harwood}, {Heesen},
  {Hoeft}, {Horellou}, {Jackson}, {Kokotanekov}, {Kondapally},
  {Kunert-Bajraszewska}, {Mahatma}, {Mahony}, {Mandal}, {McKean}, {Merloni},
  {Mingo}, {Miskolczi}, {Mooney}, {Nikiel-Wroczy{\'n}ski}, {O'Sullivan},
  {Quinn}, {Reich}, {Roskowi{\'n}ski}, {Rowlinson}, {Savini}, {Saxena},
  {Schwarz}, {Shulevski}, {Sridhar}, {Stacey}, {Urquhart}, {van der Wiel},
  {Varenius}, {Webster}, \& {Wilber}}]{shimwell19}
{Shimwell}, T.~W., {Tasse}, C., {Hardcastle}, M.~J., {et~al.} 2019, \aap, 622,
  A1

\bibitem[{{Sollerman} {et~al.}(2000){Sollerman}, {Lundqvist}, {Lindler},
  {Chevalier}, {Fransson}, {Gull}, {Pun}, \& {Sonneborn}}]{sollerman00}
{Sollerman}, J., {Lundqvist}, P., {Lindler}, D., {et~al.} 2000, \apj, 537, 861

\bibitem[{{Temim} {et~al.}(2024){Temim}, {Laming}, {Kavanagh}, {Smith},
  {Slane}, {Blair}, {De Looze}, {Bucciantini}, {Jerkstrand}, {Gountanis},
  {Sankrit}, {Milisavljevic}, {Rest}, {Lyutikov}, {DePasquale}, {Martin},
  {Drissen}, {Raymond}, {Fox}, {Modjaz}, {Spitkovsky}, \& {Strolger}}]{temim24}
{Temim}, T., {Laming}, J.~M., {Kavanagh}, P.~J., {et~al.} 2024, \apjl, 968, L18

\bibitem[{{Trimble}(1973)}]{trimble73}
{Trimble}, V. 1973, \pasp, 85, 579

\bibitem[{{van Haarlem} {et~al.}(2013){van Haarlem}, {Wise}, {Gunst}, {Heald},
  {McKean}, {Hessels}, {de Bruyn}, {Nijboer}, {Swinbank}, {Fallows},
  {Brentjens}, {Nelles}, {Beck}, {Falcke}, {Fender}, {H{\"o}randel},
  {Koopmans}, {Mann}, {Miley}, {R{\"o}ttgering}, {Stappers}, {Wijers},
  {Zaroubi}, {van den Akker}, {Alexov}, {Anderson}, {Anderson}, {van Ardenne},
  {Arts}, {Asgekar}, {Avruch}, {Batejat}, {B{\"a}hren}, {Bell}, {Bell}, {van
  Bemmel}, {Bennema}, {Bentum}, {Bernardi}, {Best}, {B{\^\i}rzan}, {Bonafede},
  {Boonstra}, {Braun}, {Bregman}, {Breitling}, {van de Brink}, {Broderick},
  {Broekema}, {Brouw}, {Br{\"u}ggen}, {Butcher}, {van Cappellen}, {Ciardi},
  {Coenen}, {Conway}, {Coolen}, {Corstanje}, {Damstra}, {Davies}, {Deller},
  {Dettmar}, {van Diepen}, {Dijkstra}, {Donker}, {Doorduin}, {Dromer}, {Drost},
  {van Duin}, {Eisl{\"o}ffel}, {van Enst}, {Ferrari}, {Frieswijk}, {Gankema},
  {Garrett}, {de Gasperin}, {Gerbers}, {de Geus}, {Grie{\ss}meier}, {Grit},
  {Gruppen}, {Hamaker}, {Hassall}, {Hoeft}, {Holties}, {Horneffer}, {van der
  Horst}, {van Houwelingen}, {Huijgen}, {Iacobelli}, {Intema}, {Jackson},
  {Jelic}, {de Jong}, {Juette}, {Kant}, {Karastergiou}, {Koers}, {Kollen},
  {Kondratiev}, {Kooistra}, {Koopman}, {Koster}, {Kuniyoshi}, {Kramer},
  {Kuper}, {Lambropoulos}, {Law}, {van Leeuwen}, {Lemaitre}, {Loose}, {Maat},
  {Macario}, {Markoff}, {Masters}, {McFadden}, {McKay-Bukowski}, {Meijering},
  {Meulman}, {Mevius}, {Middelberg}, {Millenaar}, {Miller-Jones}, {Mohan},
  {Mol}, {Morawietz}, {Morganti}, {Mulcahy}, {Mulder}, {Munk}, {Nieuwenhuis},
  {van Nieuwpoort}, {Noordam}, {Norden}, {Noutsos}, {Offringa}, {Olofsson},
  {Omar}, {Orr{\'u}}, {Overeem}, {Paas}, {Pandey-Pommier}, {Pandey}, {Pizzo},
  {Polatidis}, {Rafferty}, {Rawlings}, {Reich}, {de Reijer}, {Reitsma},
  {Renting}, {Riemers}, {Rol}, {Romein}, {Roosjen}, {Ruiter}, {Scaife}, {van
  der Schaaf}, {Scheers}, {Schellart}, {Schoenmakers}, {Schoonderbeek},
  {Serylak}, {Shulevski}, {Sluman}, {Smirnov}, {Sobey}, {Spreeuw}, {Steinmetz},
  {Sterks}, {Stiepel}, {Stuurwold}, {Tagger}, {Tang}, {Tasse}, {Thomas},
  {Thoudam}, {Toribio}, {van der Tol}, {Usov}, {van Veelen}, {van der Veen},
  {ter Veen}, {Verbiest}, {Vermeulen}, {Vermaas}, {Vocks}, {Vogt}, {de Vos},
  {van der Wal}, {van Weeren}, {Weggemans}, {Weltevrede}, {White}, {Wijnholds},
  {Wilhelmsson}, {Wucknitz}, {Yatawatta}, {Zarka}, {Zensus}, \& {van
  Zwieten}}]{vanhaarlem13}
{van Haarlem}, M.~P., {Wise}, M.~W., {Gunst}, A.~W., {et~al.} 2013, \aap, 556,
  A2

\bibitem[{{van Weeren} {et~al.}(2021){van Weeren}, {Shimwell}, {Botteon},
  {Brunetti}, {Br{\"u}ggen}, {Boxelaar}, {Cassano}, {Di Gennaro},
  {Andrade-Santos}, {Bonnassieux}, {Bonafede}, {Cuciti}, {Dallacasa}, {de
  Gasperin}, {Gastaldello}, {Hardcastle}, {Hoeft}, {Kraft}, {Mandal},
  {Rossetti}, {R{\"o}ttgering}, {Tasse}, \& {Wilber}}]{vanweeren21}
{van Weeren}, R.~J., {Shimwell}, T.~W., {Botteon}, A., {et~al.} 2021, \aap,
  651, A115

\bibitem[{{van Weeren} {et~al.}(2016){van Weeren}, Williams, Hardcastle,
  {et~al.}}]{vanweeren16}
{van Weeren}, R.~J., Williams, W.~L., Hardcastle, M.~J., {et~al.} 2016, ApJS,
  223, 2

\bibitem[{{Weisskopf} {et~al.}(2012){Weisskopf}, {Elsner}, {Kolodziejczak},
  {O'Dell}, \& {Tennant}}]{weisskopf12}
{Weisskopf}, M.~C., {Elsner}, R.~F., {Kolodziejczak}, J.~J., {O'Dell}, S.~L.,
  \& {Tennant}, A.~F. 2012, \apj, 746, 41

\bibitem[{Williams {et~al.}(2016)Williams, {van Weeren}, R{\"o}ttgering,
  {et~al.}}]{williams16}
Williams, W.~L., {van Weeren}, R.~J., R{\"o}ttgering, H. J.~A., {et~al.} 2016,
  MNRAS, 460, 2385

\end{thebibliography}

\end{document}